\documentclass[onecolumn,epsf,prl,amsmath,amssymb,12pt]{revtex4}
\usepackage{graphicx,colordvi,times}
\usepackage[usenames]{color}
\usepackage{multirow,bm}

\begin{document}
\title{Enhancement of electron-hole superfluidity in double few-layer graphene}

\author{M. Zarenia$^{1,\ast}$\email{mohammad.zarenia@uantwerpen.be}, A. Perali$^2$, D. Neilson$^{2}$, and F. M. Peeters$^1$}
\affiliation{$^1$Department of Physics, University of Antwerp,
Groenenborgerlaan 171, B-2020 Antwerpen, Belgium\\
$^2$Universit\`{a} di Camerino, 62032 Camerino (MC), Italy\\
$^\ast$E-mail:~mohammad.zarenia@uantwerpen.be}

\begin{abstract}
We propose two coupled electron-hole sheets of few-layer graphene as a  new nanostructure to observe superfluidity at enhanced densities and enhanced transition temperatures. For ABC stacked few-layer graphene we show that the strongly correlated electron-hole pairing regime is readily  accessible experimentally using current technologies.  We find for double trilayer and quadlayer graphene sheets spatially separated by a nano-thick hexagonal boron-nitride insulating barrier, that the transition temperature for electron-hole superfluidity can approach temperatures of 40 K.
\end{abstract}

%\pacs{47.37.+q,  81.05.ue, 74.20.Fg, 71.35.Lk,  74.10.+v}
\maketitle
\section*{Introduction}
The prediction of electron-hole superfluidity in spatially separated electron and hole layers has captured the attention of the scientific community  \cite{lozovik}.
The recent intense interest results from suggestions that some double-layer electron-hole systems offer the
possibility of observing a coherent superfluid state up to  temperatures approaching room temperature \cite{allen}.
Despite long standing theoretical predictions \cite{lozovik,allen,vignale} and considerable experimental efforts \cite{sivan1,sivan2,sivan3} such electron-hole superfluidity in double layered systems has not yet been observed in zero magnetic field.

Soon after the discovery of graphene \cite{novo}, a two-dimensional lattice of carbon atoms \cite{Review,beenakker}, efforts were made to look for superfluidity in graphene-based double monolayer devices \cite{zhangnew,zhangnew2}. Although the early theoretical work on graphene double monolayers  predicted room-temperature superfluidity \cite{allen}, recent Coulomb drag experiments have found no evidence of superfluidity \cite{Gorbachev2012}. It is, in fact the linear energy dispersion of monolayer graphene that makes it difficult
to access the most promising phase space region for superfluidity which is the region where the average strength of the Coulomb interactions between carriers is much larger than their average kinetic energy.
The reason is the following.  The most favourable conditions for the electron-hole pairing are achieved at small interlayer separations $d$, when $k_Fd \ll 1$.
In this optimal limit  the behavior of the system is determined by the dimensionless interaction parameter $r_s=\langle V\rangle/E_F$ \cite{lozovic2}. $E_F$ is the Fermi energy and $\langle V\rangle=e^2/(\kappa\langle r_0\rangle)$ is the average Coulomb energy for the mean inter-particle spacing in a sheet, $\langle r_0\rangle=1/\sqrt{n\pi}$, $n$ is the charge carrier density in the sheet and $\kappa$ is the dielectric constant of the barrier.
For monolayer graphene, $E_F=\hbar v_F k_F$, where the graphene Fermi velocity $v_F\simeq 10^6$ ms$^{-1}$,  the Fermi momentum $k_F=\sqrt{4\pi n/g_sg_v}$, and the spin (valley) degeneracy for graphene is $g_s=2$ ($g_v=2$).  This gives for monolayer graphene a value of $r_s=e^2/[\kappa \hbar  v_F]$ that is constant, independent of the density \cite{sarma}. The dielectric constant for a hexagonal boron-nitride (h-BN) insulating barrier is $\kappa\approx 3$ \cite{Dean}, giving $r_s$ a very small (and fixed) value of only $r_s=0.7$. Calculations for double monolayer graphene unfortunately indicate that unless the parameter $r_s$  exceeds $r_s\agt 2.3$, screening of the electron-hole attractive interaction suppresses superfluidity at all practicable non-zero temperatures \cite{lozovic2}.  This makes it very difficult to experimentally realize electron-hole superfluidity in double monolayer graphene \cite{Gorbachev2012}.

Recently it has been suggested that a pair of bilayer graphene sheets is a promising system for observing high temperature superfluidity \cite{david}. In contrast with monolayer graphene, bilayer graphene has a quadratic dispersion at low energies so its parameter $r_s$  has an inverse density dependence $r_s\sim 1/\sqrt{n}$ that is familiar from metals and semiconductors. This density dependence makes it possible to experimentally access the strongly interacting regime at large $r_s$ simply by reducing $n$. Calculations for double bilayer graphene indicate that the interaction parameter $r_s$  must exceed a value similar to that for double monolayer graphene, $r_s \agt 2.3$, in order for the superfluid to condense at non-zero temperatures \cite{david}. It should be noted that outside the optimal region for superfluidity $k_F d \ll 1$, superfluid state properties start to be sensitive also to the barrier thickness $d$.  With increasing  $d>1/k_F$, there is ({\it i}) an increase in the minimum value of $r_s$ at which finite-temperature superfluidity occurs, and ({\it ii}) a decrease in the maximum gap $\Delta_{\mathrm{max}}$.
\begin{figure}[!h]
\centering \vspace{0 mm}
\includegraphics[width=0.8\textwidth]{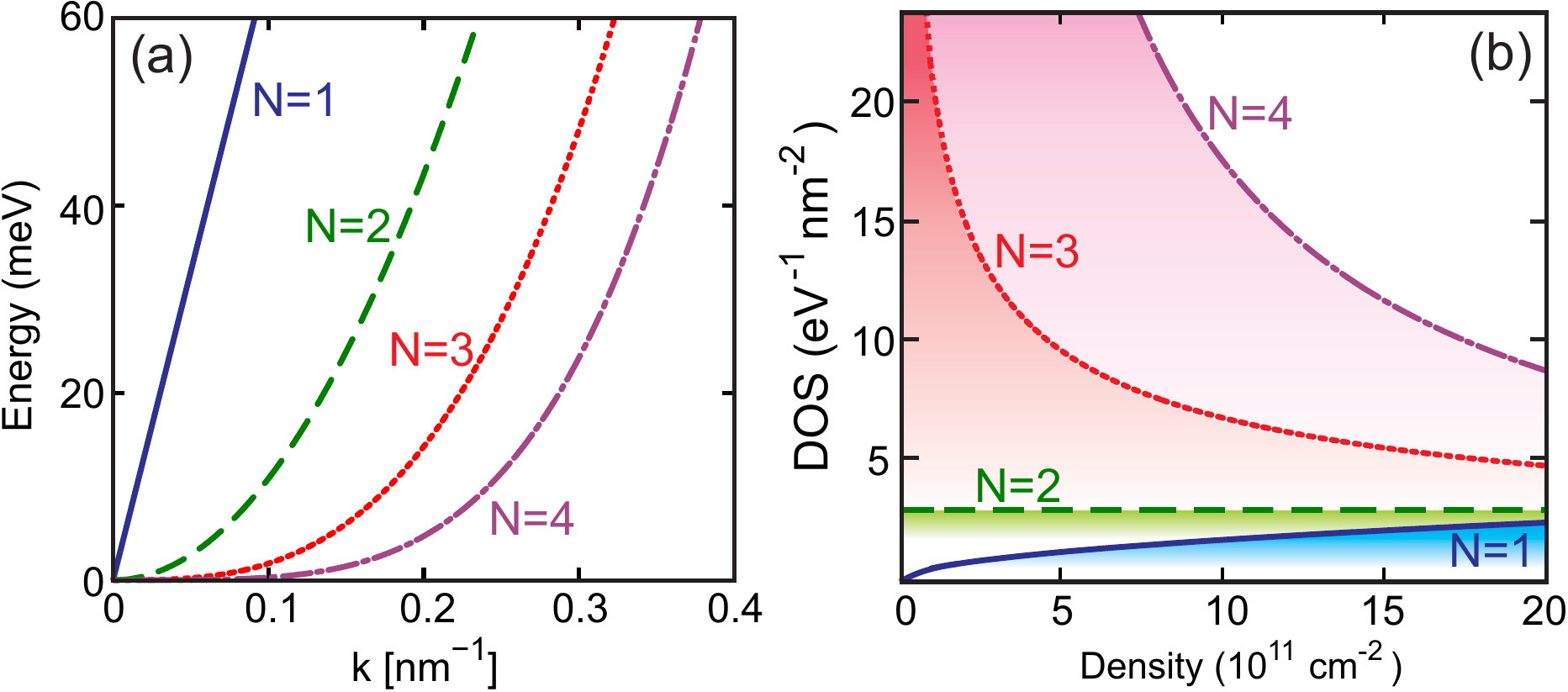}
\caption{(Color online) (a) Lowest positive energy band in monolayer ($N=1$), bilayer ($N=2$), trilayer ($N=3$), and quadlayer ($N=4$) graphene.
(b) The density of states at the Fermi energy for $N=1$ to $N=4$ as function of the carrier density.}
\label{fig1}
\end{figure}

In this paper we investigate the possibility of using graphene systems consisting of  double few-layers in excess of two (that is, the bilayer case) in order to access regions of phase space that are even more strongly interacting, with very large values of $r_s$.  We shall see that such systems offer further potential advantages arising from divergences in the density of states caused by van Hove singularities.

Based on the two-band Dirac-Weyl equation describing the lowest energy band in ABC stacked $N$-layer graphene,  the  energy dispersion of the conduction band is given by \cite{min,katsnelson}
\begin{equation}
E^{(N)}(k)= \left\{(\hbar v_F)^{N}/t^{N-1}\right\} k^{N}\ ,
\end{equation}
where  $t\approx 400$ meV is the interlayer hopping term in few-layer graphene.  Figure \ref{fig1}(a) shows $E^{(N)}(k)$ for $N=1$ to $4$.
We then obtain in $N$-layer graphene,
\begin{equation}
r_s=\left\{ \frac{e^2t^{N-1}}{\kappa (\hbar v_F)^N \sqrt{\pi^{N-1}}}\right\} \frac{1}{n^{(N-1)/2}}\ .
\end{equation}
(Note this expression for $r_s$ reduces to the ratio of $r_0$ to the effective Bohr radius, only in the case of quadratic bands with  $g_s=2$ and $g_v=1$.)

Table I compares the values of $r_s$ for the typical electron densities found in graphene sheets for $N$-layer graphene, with $N$ ranging from $N=1$ (monolayer), to $N=4$ (quadlayer).  The table shows that few-layer graphene offers dramatic opportunities for producing extremely strongly interacting systems at experimentally accessible densities.
%

%%%%%%%%%%%%%%%%%%%%%%%%%%%%%%%%%%%%%%%%%%%%%%%%%%%%%%%%%%%%%
\begin{table}[!h]
\caption{Values of the parameter $r_s$ for few-layer graphene.}
\begin{tabular}{c c c c c}
\hline\hline
Density (cm$^{-2}$) & monolayer   & bilayer & trilayer &  quadlayer     \\
\hline
$5\times 10^{12}$    &  $0.7$          & $1$      & $2$     &  $3$               \\
%\hline
$1\times 10^{12}$    &  $0.7$          & $3$      &  $8$    &  $29$              \\
%\hline
$5\times 10^{11}$    &  $0.7$          & $4$      &  $17$   & $83$              \\
%\hline
$1\times 10^{11}$    &  $0.7$          & $8$      &  $86$   & $930$             \\
\hline
\hline
 \end{tabular}\
\label{table1}
\end{table}
%%%%%%%%%%%%%%%%%%%%%%%%%%%%%%%%%%%%%%%%%%%%%%%%%%%%%%%%%%%%%
The ability to access large $r_s$ values in few-layer graphene, and thus to reach the strong electron-hole pairing regime in an experimentally accessible range of densities, motivates us to propose few-layer graphene as a system to observe electron-hole superfluidity at enhanced densities and transition temperatures.

Experimental realization of few-layer graphene is readily within the grasp of current technology since few-layer graphene sheets can be fabricated in large areas by both mechanical exfoliation \cite{Ferrari2006,Zhang2005} and by chemical techniques \cite{Berger2004,Shih2011,Mahanandia2014} from graphite with controlled stacking order. References \cite{craciun,bao,mak} are examples of experimental studies on electronic and transport properties in trilayer graphene.
\begin{figure}[!h]
\centering \vspace{-2 mm}
\includegraphics[width=0.8\textwidth]{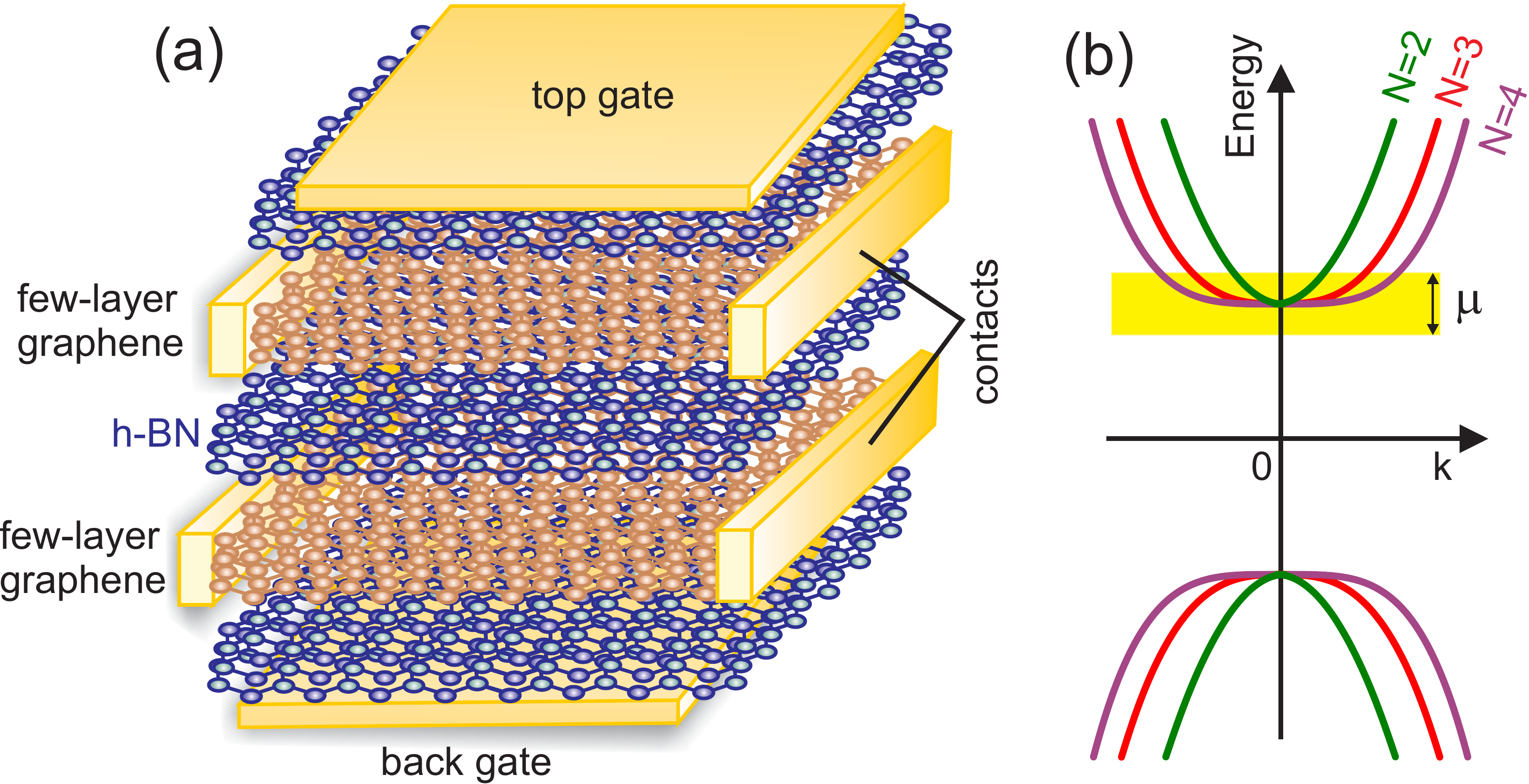}
\caption{(Color online) (a) Schematic illustration of two few-layer graphene sheets separated by a thin barrier of h-BN layers. The electrons and holes are induced in the separately electrically contacted upper and lower graphene sheets by top and back gates. (b) Sketch of the energy bands of gapped bilayer $N=2$, trilayer $N=3$, and quadlayer $N=4$ graphene. The yellow region around the bottom of the conduction band indicates the range of values of the chemical potential $\mu$ for our range of carrier densities.}\label{fig2}
\end{figure}
%`

A schematic setup of our proposed system is depicted in Fig.\ \ref{fig2}(a).  There are two parallel few-layer graphene sheets. The upper sheet of electrons and the lower sheet of holes are influenced by the top and back gates. The two sheets are separated by a thin h-BN insulating barrier to prevent tunneling  between the sheets and electron-hole recombination.  The separation of the graphene
sheets can be as small  as 1 nm (three h-BN layers) and still provide a potential barrier high enough to suppress
tunneling \cite{Gorbachev2012,Geim2013}.  As well as the top and bottom metal gates there are separate electrical contacts
to the two sheets, allowing independent control over the carrier density in each sheet.

Our aim is to provide experimental indicators for system design for observing high-T$_\mathrm{c}$ electron-hole superfluidity for the first time.
We first evaluate the superfluid energy gap within an extended mean-field approach.
Inducing electrons or holes in a few-layer graphene sheet using metallic gates imposes a perpendicular electric field.  Experiments show
that such a perpendicular electric field induces a band gap in the single-particle spectrum in bilayer \cite{Zhang2009} and trilayer \cite{craciun,bao,Lui2011,Zou2013} graphene band structures, and
recent theoretical studies predict a similar effect in the energy spectrum of all few-layer graphene \cite{Avetisyan,Tang2011}.
This induced band gap makes the contributions from the graphene valence band small, and for the calculation of the superfluid gap we need to consider only contributions from the conduction band. A sketch of the energy bands for gapped bilayer, trilayer and quadlayer graphene is shown in Fig. \ref{fig2}(b). The highlighted region indicates the range of values for the chemical potential $\mu$ corresponding to the range of carrier densities considered here.
\section*{Methods}
We fix the electron and hole chemical potentials $\mu$ and densities $n$ to be equal. The
equations for $\Delta_\mathbf{k}$ and $\mu$ are
\begin{equation}
\label{Deltaeqn}
  \Delta_\mathbf{k} = -\sum_{\mathbf{k'}}
F_{\mathbf{kk'}} V_{\mathbf{k}-\mathbf{k'}}\frac{\Delta_{\mathbf{k'}}}{2E_{k'}}\ ;
~~~~~~~~~
n=g_sg_v \sum_{\mathbf{k}}v_{k}^2,
\end{equation}
$\Delta_k$ is the wave-vector dependent zero temperature gap generated from pairing of electrons and holes in the conduction band. $E_{k}=\sqrt{{\epsilon_{k}}{^2}+{\Delta_{k}}{^2}}$ and  $v_{k}^2=({1}/{2})\big(1-{\epsilon_k}/{E_{k}}\big)$, with $\epsilon_{k}=  E^{(N)}(k)-\mu$. The factor $F_{\mathbf{kk'}}=\big[1+\cos [N(\phi_k-\phi_{k'})]\big]/2$ for $N$-layer graphene is
associated with the square of the overlap between the single-particle states $|k\rangle$ and $|k'\rangle$ \cite{lozovic2}.

We can take into account the finite thickness of each $N$-layer graphene sheet using an effective barrier thickness set equal to the physical distance perpendicular to the interface between the midpoints of the two $N$-layer sheets. We find this is a good approximation provided that the sheet's physical thickness is less than the thickness of the barrier separating the two $N$-layer sheets. This simplification is possible thanks to the strong hybridization of the electron states between the $N$-layers within a sheet \cite{Lui2011,Zou2013,Avetisyan}. To check this, we compared the electron-hole Coulomb pairing interaction for the hybridized electrons and holes in the case of double bilayer sheets of graphene, to this approximation of two thin sheets plus effective barrier thickness. We found for physical barrier thicknesses as small as the thickness of the bilayer sheet ($\approx 0.3$ nm), that the resulting shift in the Coulomb pairing interaction did not exceed 5\%. For the remainder of the paper we will denote by $d$ the effective barrier thickness. A construction similar to this is used with coupled double electron-hole quantum wells in GaAs where the finite width of the quantum wells can be treated as a form factor multiplying the electron-hole Coulomb interaction \cite{Tan}.

We evaluate $V_{\mathbf{k}-\mathbf{k'}}$, the static screened Coulomb interaction between electrons and holes in the two $N$-layer sheets \cite{david,david2,screening,screening1}, starting from the random phase approximation (RPA) in the normal state,
\begin{eqnarray}
\!\!V_{\mathbf{q}}\!\!&=&\!\! \frac{v(q)\exp(-qd)}{1+2v(q)\Pi_0(q)+v(q)^2\Pi_0(q)^2[1-\exp(-2qd)]} \label{vq}\\
              &\simeq&\frac{v(q)\exp(-qd)}{1+2v(q)\Pi_0(q)}\label{vqapp}
\end{eqnarray}
where $v(q)= -2\pi e^2/[\kappa q]$ is the attractive bare Coulomb interaction for $d=0$ and $\Pi_0(q)$ the normal state particle-hole polarization bubble. The approximate expression in Eq.\ (\ref{vqapp}) uses the property that the most favourable conditions for pairing will occur when $k_Fd\ll 1$, which is the case over the low-density range in which we work. For example, for $d=2$ nm and density $n=6.5\times 10^{11}$ cm$^2$, which is the highest onset density for superfluidity found in our calculations, $k_Fd=0.28$.

For small momentum exchange, we can extend Eq.\ (\ref{vqapp}) to the broken symmetry phase at $T=0$ by writing the full polarization as $\Pi_0(q)=\Pi_0^{(n)}(q)+\Pi_0^{(a)}(q)$, where $\Pi_0^{(n)}(q)$ and $\Pi_0^{(a)}(q)$ are the normal and anomalous polarization bubbles in the superfluid state.  At zero-temperature,
\begin{eqnarray}\label{P}
  \Pi_{0}^{(n)}(q) &=& -2g_v\sum_{\mathbf{k}} F_{\mathbf{kk-q}}\frac{{u_k}^2 {v_{\mathbf{k}-\mathbf{q}}}^2+{v_k}^2 {u_{\mathbf{k}-\mathbf{q}}}^2}{E_{k}+E_{\mathbf{k}-\mathbf{q}}}\ ,~~~~~~\nonumber \\
  \Pi_{0}^{(a)}(q) &=& 2g_v\sum_{\mathbf{k}} F_{\mathbf{kk-q}}\frac{2u_k u_{\mathbf{k}-\mathbf{q}}v_{k}v_{\mathbf{k}-\mathbf{q}}}{E_{k}+E_{\mathbf{k}-\mathbf{q}}}\ ,
\end{eqnarray}
where $u_{k}^2=\big(1+{\epsilon_k}/{E_{k}}\big)/2$.

$\Pi_0^{(n)}(q)$ and $\Pi_0^{(a)}(q)$ are numerically calculated self-consistently for the few-layer graphene in the superfluid state \cite{lozovic2,david}. This procedure follows the approach of Ref. \cite{lozovic2}.  For $k_Fd\ll 1$, the full RPA-BCS screened interaction reduces to the present approximation (Eqs. (\ref{vqapp}) and (\ref{P})).
Results from this approach have been tested successfully against Diffusion Quantum Monte Carlo (DQMC) results for electron-hole double layer systems \cite{needs,needs2,david2}. Reference \cite{david2} found that RPA screening in the superfluid state gives satisfactory agreement with the condensate fractions $c=\sum_{\mathbf{k}}u_{k}^2v_{k}^2/\sum_{\mathbf{k}}v_{k}^2$ \cite{salasnich} calculated within DQMC over a wide parameter range, demonstrating that using self-consistent  screening within the superfluid state is a good mean-field approximation.

We can neglect the intralayer correlations between electrons in the same sheet for two reasons. At high densities the electron-electron interactions are weak, while at low densities the compact pairs are weakly interacting.  The satisfactory agreement for $r_s\alt 3$ between the DQMC approach in Ref. \cite{needs,needs2} and our present mean-field approach confirm that the intralayer correlations have little effect on the superfluid properties for $r_s\alt 3$, consistent with the conclusion drawn by comparing the gaps reported in Fig.\ 2 of Ref.\ \cite{Zhu95}, which included intralayer correlations, with the gaps calculated in Ref.\ \cite{Pieri}, which neglected these correlations.  This comparison shows, at most, a $10$-$20$\% effect on the zero temperature gap.

In the superfluid state at low temperature, $\Pi_0(q)$ is suppressed at small momenta $q$ because of the opening of the superfluid energy gap $\Delta$ at the Fermi surface \cite{GortelSwier96,BMSMcomment}. The suppression of screening permits Cooper pairs to form. However for both
double monolayer and double bilayer graphene, unless the values of the interaction parameter $r_s$ exceeds $\sim2.3$,  the screening remains too strong for superfluidity to occur at any practicable zero temperature \cite{lozovic2,david}. When $r_s\agt 2.3$, the screening has become sufficiently suppressed by the opening of the superfluid gap that the self-consistent Cooper pairing can be strong. At finite temperature, the underlying physics leading to the superfluid transition is that for strong electron-hole pairing the fluctuations of the order parameter determine a (pseudo)gap at the critical temperature of the same order as the zero temperature superfluid gap. This large pseudogap should lead to a suppression of the screening similar to the suppression at zero temperature caused by the superfluid gap.   A large pseudogap of the same order as the $T=0$ superfluid gap has been experimentally observed and theoretically investigated in ultracold fermionic gases in both three-dimensional \cite{GSDJPPS2010} and two-dimensional traps \cite{FeldMarsiglio,FeldMarsiglio2}.
\section*{Results}
As a consequence of the different energy dispersions $E^{(N)}(k)$, the energy dependence of the density of states (DOS) changes dramatically with the number of layers $N$,
\begin{equation}
DOS^{(N)}(E)=  \frac{2\pi}{N}\frac{t^{2(N-1)/N}}{(\hbar v_F)^2}E^{(2/N)-1}.
\end{equation}
Figure \ref{fig1}(b) shows the dependence of $DOS^{(N)}(E_F)$  at the Fermi energy on carrier density $n$.  For monolayer graphene $DOS^{(1)}(E_F)$ depends linearly on $n$, for bilayer graphene $DOS^{(2)}(E_F)$ is a constant. For trilayer and quadlayer graphene, $DOS^{(N)}(E_F)$ decreases with $n$. $DOS^{(3)}(E_F)$ and $DOS^{(4)}(E_F)$ for small densities are much larger than $DOS^{(1)}(E_F)$ and $DOS^{(2)}(E_F)$ because of their van Hove singularities at the band bottom. At very high densities, lying far above our present range of interest, the $DOS^{(3)}(E_F)$ and $DOS^{(4)}(E_F)$ become smaller than $DOS^{(1)}(E_F)$ and $DOS^{(2)}(E_F)$ .

\begin{figure}[!h]
\centering
\includegraphics[width=0.65\textwidth] {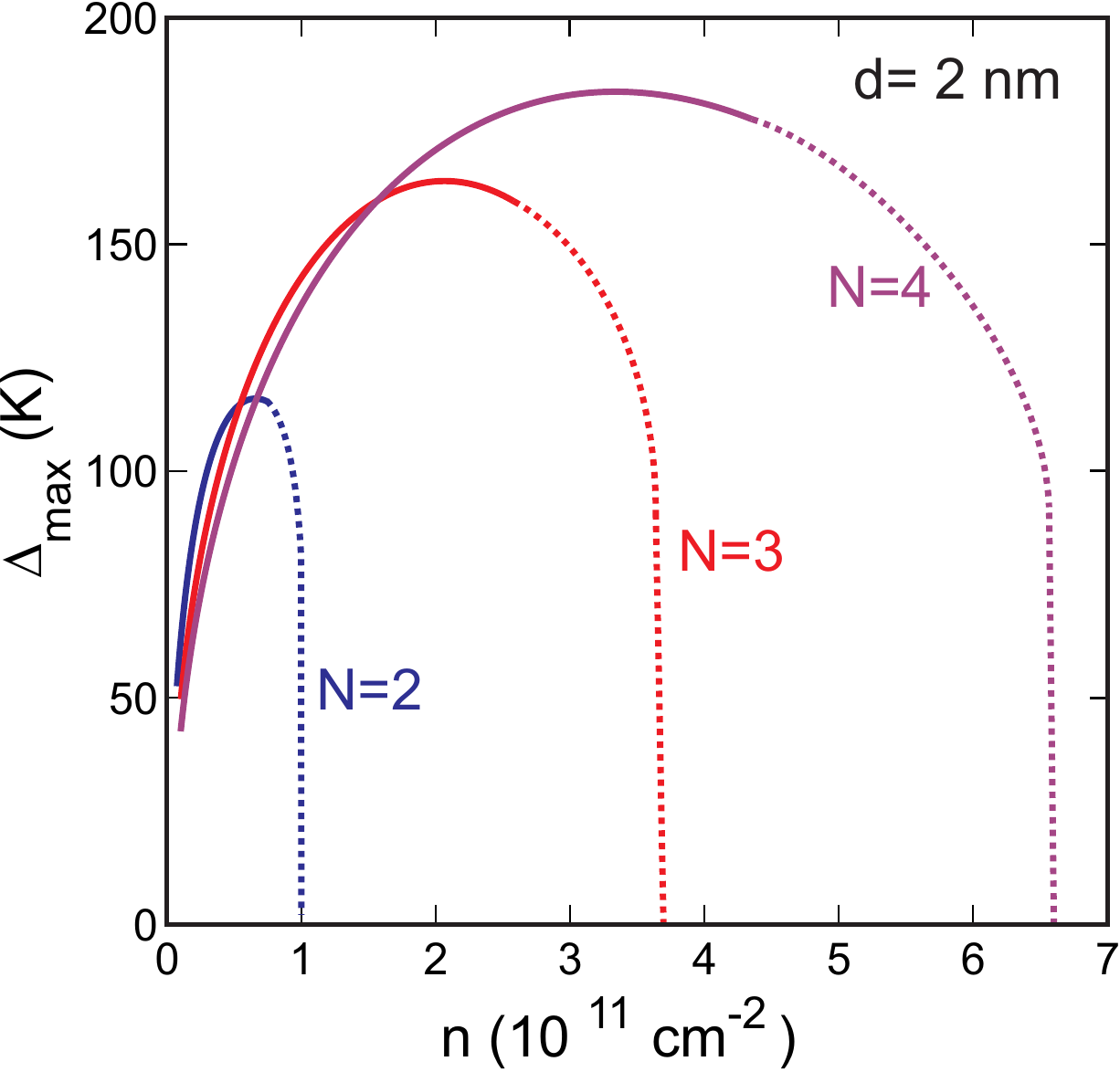}
\vspace{0 mm}
 \caption{(Color online) Maximum of superfluid gap at zero temperature in double bilayer $N=2$, double trilayer $N=3$, and double quadlayer $N=4$ graphene  for effective barrier thickness $d=2$ nm.
Solid lines: BEC regime;  Dotted lines: BEC-BCS crossover regime. The BCS regime is not reached.}\label{fig3}
\end{figure}
We self-consistently solve Eqs. (\ref{Deltaeqn}), (\ref{vqapp}) and (\ref{P}) for the momentum-dependent gap $\Delta_{\bf k}$. Figure \ref{fig3} shows $\Delta_{\mathrm{max}}$, the maximum $\Delta_{\bf k}$ for coupled $N$-layer graphene with $N=2$, $3$ and $4$ for effective barrier thickness $d=2$ nm.
For densities above an onset density $n_c$, if there is any superfluidity at all, the gap would be extremely small, $\Delta_{\mathrm{max}}\ll\! 1$ K.
At the onset density there is a sudden discontinuous jump in $\Delta_{\mathrm{max}}$ to high energies of the order of the chemical potential $\mu$. For $N=3$ and $4$, the pairing interactions are stronger as compared to $N=2$, and the large-gap superfluidity is seen to persist up to significantly higher $n_c$.

We find that the peak in $\Delta_{\mathrm{max}}$ is located in the BEC regime. Figure \ref{fig3} shows the BEC and BEC-BCS crossover regimes which we determine using the following criterion. Condensate fractions $c>0.8$ correspond to the BEC regime of compact electron-hole pairs on the scale of $r_0$ and $0.2<c\leq 0.8$ correspond to the BEC-BCS crossover regime \cite{david2}.  $c\leq 0.2$  would correspond to the weak-coupled BCS regime, but screening suppresses superfluidity at densities $n>n_c$  before the BCS regime can be reached. In the BEC regime the effect of screening is much less dramatic. We find that screening reduces the gap in the BEC regime by a factor of two or less compared with the corresponding gap calculated without screening. The reason is the compact nature of the electron-hole pairs in the BEC regime. The relation between the density dependence of the gap and the average effective Coulomb interaction and polarization function are discussed in the supplementary information (see Fig. S1).
\begin{figure}[!h]
\centering
\includegraphics[width=0.8\textwidth] {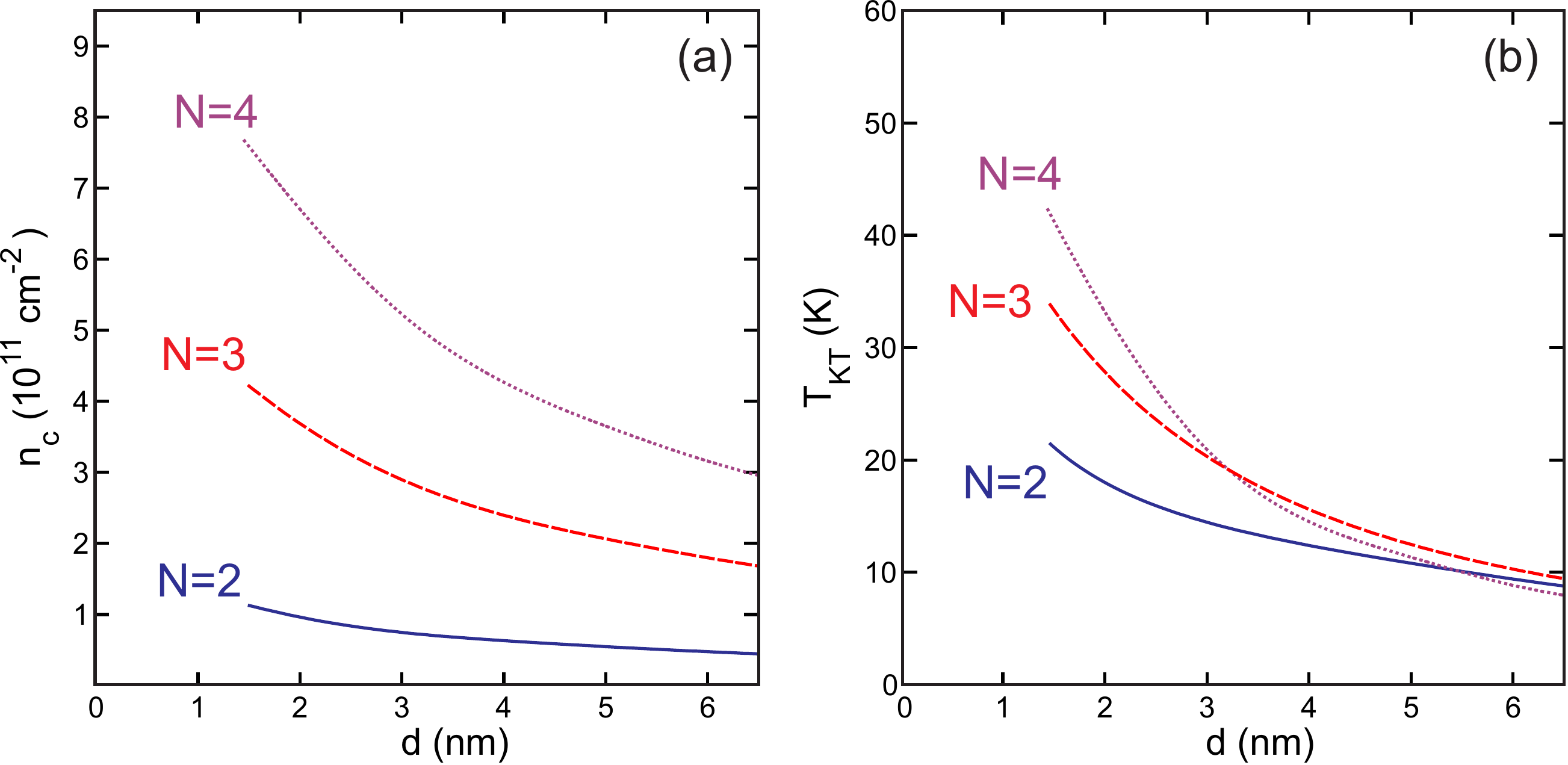}
\vspace{0 mm}
 \caption{(Color online) (a) Maximum densities $n_c$ for finite temperature superfluidity for the $N=2$, $3$, and $4$ few-layer graphene systems,
as function of effective barrier thickness $d$.
 (b) Corresponding maximum  Kosterlitz-Thouless transition temperatures $T_{KT}$ as function of $d$.}\label{fig4}
\end{figure}

We next determine the variation of the onset density on the number of layers $N$.  Figure \ref{fig4}(a) shows  the different $n_c$  as functions of $d$.  At a fixed $d$, the onset density increases with the number of layers $N$, and the differences become very significant as $d$ decreases.  For the double quadlayer system, $n_c$ approaches $10^{12}$ cm$^{-2}$ for $d=1.5$ nm.

A central  concern for experiments  and device applications is the predicted temperature $T_c$ for the superfluid phase transition.  While in two dimensions $T_c$ is not linearly related to the value of the zero temperature gap $\Delta$ \cite{KT1973},  nevertheless a large value of $\Delta$ through strong pairing is an essential prerequisite for a high $T_c$. For superfluids in two-dimensions, an upper bound on the transition temperature is the Kosterlitz-Thouless temperature $T_{KT}$ \cite{KT1973}.  This is determined from
\begin{equation}
T_{KT} = (\pi/2)J(T_{KT}),
\label{TKT}
\end{equation}
where $J(T)$ denotes the superfluid stiffness (the average kinetic energy of the Cooper pairs). At zero $T$,  $J(0)$ is proportional to the superfluid density $J(0)=\rho_s(0)/2$, since $\rho_s(0)$ controls the phase stiffness of the complex superfluid order parameter.
In the mean field approach,  $\rho_s(T)$ falls off only slowly with temperature when $T$ is small compared to the zero-temperature gap so $\rho_s(T)$ is well approximated by $\rho_s(0)$ when $k_BT\ll \Delta(0)$. In an isotropic system,
$\rho_s(0)$ at the mean field level is determined from \cite{benfatto,benfatto2}
\begin{equation}\label{rhos}
\rho_s(0) =g_sg_v\sum_{\mathbf{k}} \left[1/m^\star(k)\right] v_{k}^2,
\end{equation}
where $1/m^\star(k)={\partial^2\epsilon_k}/{\partial k^2}$  is momentum dependent.
For bilayer graphene $\rho_s(0)=[2(\hbar v_F)^2/t]n$.   Note  it is only for $N=2$ that $\rho_s(0)$ is proportional to $n$.

Figure \ref{fig4}(b) shows the maximum $T_{KT}$. This occurs at the onset density $n_c$. The maximum $T_{KT}$ is plotted as a function of  $d$ for $N=2$, $3$, and $4$.   Results are shown only for cases $k_BT_{KT} < 0.5\Delta_{\mathrm{max}}$.
We see that for the same $d$, the enhancement of the $T_{KT}$ with increasing $N$ is significantly less  than the corresponding enhancement of the onset density $n_c$ (Fig.\  \ref{fig4}(a)).
The reason is that in the relevant density range, the effective mass $m^\star(k)$ for $N=3$  and $N=4$ is  larger than the effective mass $m^\star$ for $N=2$.  This has the effect of reducing the $T=0$ superfluid density for $N>2$ as compared with the $N=2$ case.  The larger $m^\star(k)$ arise from the different band curvatures, and they partially compensate  both amplification effects in the pairing gap and also the increased suppression of  Coulomb screening caused by the van Hove singularities present in the DOS for $N=3$ and $N=4$. If we increase the number of layers above $N=4$, we expect these compensating effects will increase, making further net gains in $T_c$ less significant.

We see in Fig.\  \ref{fig4}(b) that changing from an $N=2$ to an $N=4$ sample with the same $d\leq 2$ nm, has the effect of doubling $T_{KT}$.  Therefore in double quadlayer graphene heterostructures at  currently experimentally attainable  densities, $\sim 10^{12}$ cm$^{-2}$, transition temperatures  can approach temperatures of the order of 40 K.
This strongly suggests that  electron-hole superfluidity leading to counter-flow superconductivity should be readily detectable in such samples using current technologies.

The superfluid transition should persist to higher temperatures than possible coherent states in one of the graphene sheets. This is because the superfluidity is driven by the attractive electron-hole interaction between layers which will dominate over the repulsive electron-electron interactions within a layer. Furthermore, with small effective barrier thicknesses $d\sim2$ nm, the mean electron-hole spacing in the pairs is much smaller than $r_0$ for the range of densities we are considering. Thus the electron-hole pairing interaction will be much stronger than the corresponding electron-electron repulsion.

The superfluid phenomena that we have been considering are not affected by disorder for the following reasons.  Reference \cite{MacD} states that disorder will not destroy superfluidity for sufficiently low impurity concentrations $n_i$ satisfying the condition $n_i\pi d^2 < k_Fd$.  With graphene-hBN interfaces, $n_i$ can be $\sim10^{10}$ cm$^{-2}$ \cite{yacoby,crommie,xue}, so that even for wide barriers, $d=10$ nm, the value of $n_i\pi d^2\leq 0.1$.  At the onset densities for superfluidity we are already in the crossover regime, resulting in values of  $k_Fd \lesssim 1$.  Thus all the samples we consider are well within the condition for negligible effects of disorder specified in Ref. \cite{MacD}.  Furthermore, even at the highest densities for which finite-temperature superfluidity occurs, we are already in the crossover regime with large superfluid gaps $\Delta_{\mathrm{max}} > 10$ meV.  Abergel {\it et al.} \cite{Abergel1,Abergel2} find that for superfluid gaps greater than a few meV, the level of fluctuations found in h-BN substrates is insufficient to destroy the electron-hole superfluidity.  (It is also interesting to note that Efimkin {\it et al.}\cite{efimkin} quote a minimum superfluid transition temperature of 19.8 K in the weakly-interacting BCS limit in the related system of coupled electron-hole graphene monolayers even with a disorder concentration which is an order of magnitude greater than our $\sim10^{10}$ cm$^{-2}$.) Finally, we recall that the fluctuations of the chemical potential $\mu$ associated with disorder \cite{Abergel1,Abergel2}  are in one-to-one correspondence with density fluctuations.  In the crossover regime, Ref. \cite{Pieri} showed that superfluid properties are insensitive to imbalances in the electron and hole densities less than 30\%.  This is because the sizeable smearing of the Fermi surfaces means that perfect matching of the Fermi surfaces is not necessary to stabilize superfluidity in the crossover regime.

In summary, we predict  enhanced electron-hole superfluidity at temperatures up to $\sim 40$ K in double few-layer sheets of graphene with large carrier densities as high as $10^{12}$ cm$^{-2}$.
An important element of the physical mechanism is that increasing the number of graphene layers in each sheet has the effect of greatly enhancing the density of states (Fig.\ \ref{fig1}(b)).  This enhancement projects the  sheets into the strongly interacting regime, leading to strong electron-hole pairing  at large accessible densities. Over the full range of system parameters considered, we established that disorder effects will play a minor role on superfluid properties. The experimental parameters of our proposed device have all been attained in related graphene systems \cite{Lui2011,Zou2013,Tutuc}.
\ \\

\section*{Acknowledgements}
We thank L. Benfatto, S. De Palo, and G. Senatore for  helpful comments. This work was partially supported by the Flemish Science Foundation (FWO-Vl) and the European Science Foundation (POLATOM).

\newpage
\section*{SUPPLEMENTARY INFORMATION}
In Fig. \ref{figs1}(a) we show the effective electron-hole Coulomb interaction, including self-consistent screening in the superfluid phase, which has been averaged over the wave-vector $q$ transferred in the pairing process. We present our results for effective barrier thickness $d=2$ nm. Confirming our results in Fig. 3, the average Coulomb interaction determines the onset density $n_c$ of the superfluid gap. Near $n_c$ it exhibits a steep density dependence because of the strong suppression of the pairing due to the Coulomb screening (as shown in Figs. \ref{figs1}(b,c,d)). The non-monotonic behavior of the gap as a function of density, seen in Fig. 3, is not related to the density dependence of the effective interaction or the Coulomb screening. It is a known intrinsic consequence of the BCS-BEC crossover phenomenon for pairing mediated by Coulomb attraction. Indeed this feature is also present in the unscreened case (see Refs. \cite{david,Pieri}).
\setcounter{figure}{0}
\makeatletter
\renewcommand{\thefigure}{S\@arabic\c@figure}
\makeatother
\begin{figure}[!h]
\centering \vspace{-1 mm}
\includegraphics[width=0.9\textwidth] {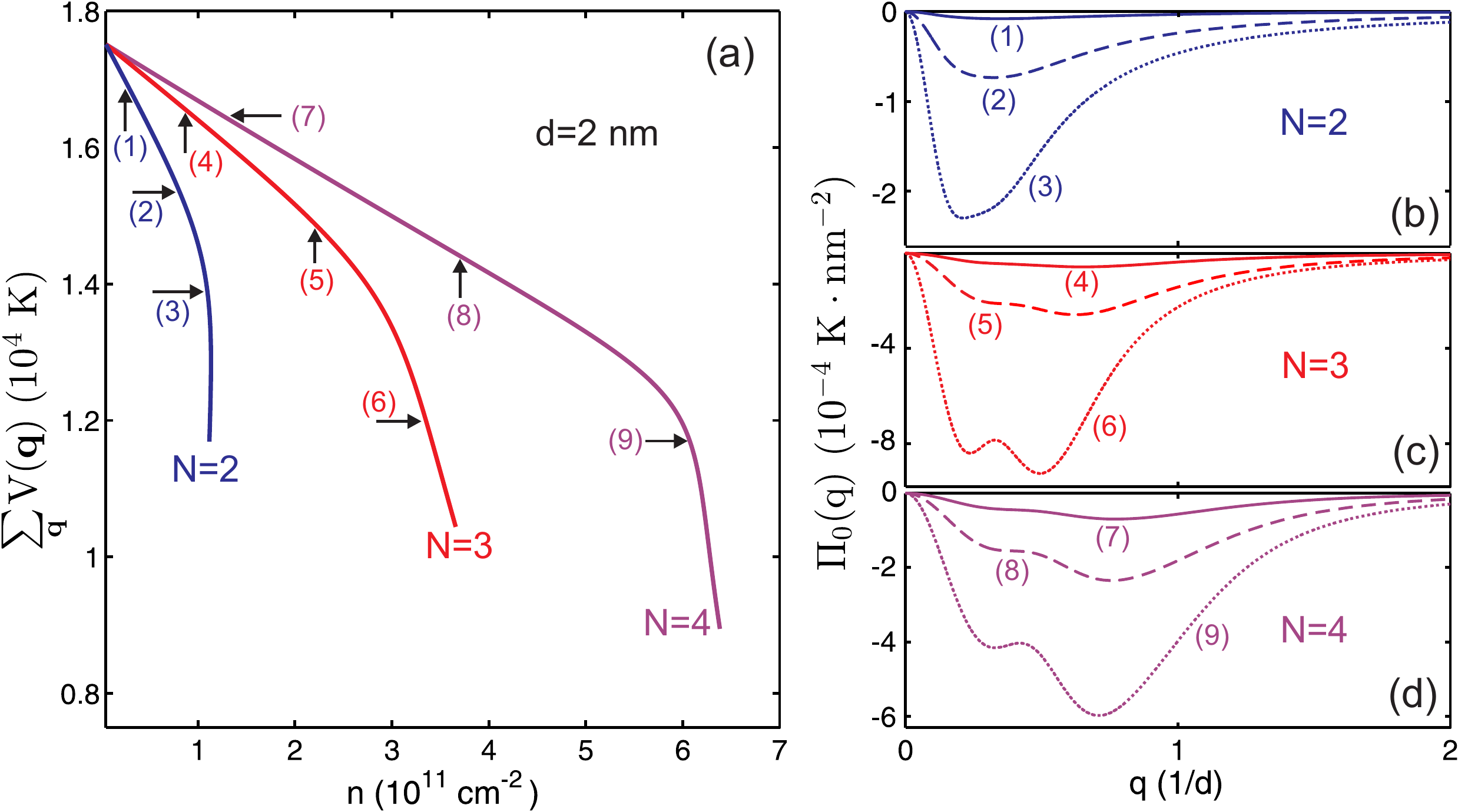}
\caption{(Color online) (a) Electron-hole Coulomb interaction averaged over momentum transfer $q$ as a function of carrier density for effective barrier thickness $d=2$ nm. Right panels: Full polarization function for (b) $N=2$, (c) $N=3$, and (d) $N=4$ systems. The numbers label the corresponding densities in Fig. \ref{figs1}(a).}
\label{figs1}
\end{figure}

Figures \ref{figs1}(b,c,d) show in detail the wave-vector dependence of the total screening bubble for different number of layers $N$ and for three characteristic densities as labelled in Fig. \ref{figs1}(a). In all cases the screening bubble is strongly suppressed by the superfluid gap opening at small wave-vectors. It is the small wave-vector contributions that play the most important part for electron-hole pairing by the Coulomb interaction. As expected, for large wave-vectors the screening bubble is not strongly affected by the opening of the superfluid gap. For small densities in the BEC regime, the chemical potential becomes negative and the screening bubble is strongly suppressed everywhere. It is only close to the onset density that the screening bubble becomes sufficiently large for it to kill the superfluidity for all the $N$ we have considered.
For $N>2$, panels \ref{figs1}(c) and \ref{figs1}(d), the absolute values of the screening bubble increase for all wave-vectors due to the enhanced density of states (DOS) for $N=3$ and $N=4$. In addition to the screening, the enhancement of the DOS affects the gap equation. There is an amplification of the electron-hole pairing, leading to the enhancement of the superfluid gap seen in Fig. 3 in the main text.

\end{document}